\documentstyle[11pt,emulateapj]{article}
\def\p{\partial}

\begin{document}
\submitted{Received 1999 March 2 ; accepted 1999 April 26}

\title
    {Standard Solar Models in the Light of New Helioseismic Constraints\\    
II. Mixing below the Convective Zone}
    
\author{A. S. BRUN$^{1,2}$, S. TURCK-CHIEZE}

\affil{$^1$DSM/DAPNIA/Service d'Astrophysique, CEA Saclay,
 91191 Gif-sur-Yvette, France\\}
 
\and
 
\author{J. P. ZAHN}

\affil{$^2$D\'epartement d'Astrophysique Stellaire et Galactique, Observatoire de 
Paris, Section Meudon, 92195 Meudon, France\\}

\begin{abstract}

In previous work, we have shown that recent updated standard solar models  
cannot reproduce the radial profile of the sound speed at the base of the 
convective zone and fail to predict the photospheric lithium abundance. 
In parallel, helioseismology has shown that the transition from differential rotation in the 
convective zone to almost uniform rotation in the radiative solar interior occurs in a shallow 
layer called the tachocline. This layer is presumably the seat of a large scale circulation and 
of turbulent motions. Here, we introduce a macroscopic transport term in the structure equations, 
which is based on a 
hydrodynamical description of the tachocline proposed by Spiegel \& Zahn, and we calculate the mixing induced within this layer. We discuss the influence of different parameters that represent the tachocline 
thickness, the Brunt-V\"aiss\"al\"a frequency at the base of the convective zone, and the time 
dependence of this mixing process along the Sun's evolution. We show that the 
introduction of such a process inhibits the microscopic 
diffusion by about 25 \%. Starting from models including a pre-main sequence evolution, we obtain: 

a) a good agreement with observed photospheric chemical abundance of light elements such as 
$^3$He, $^4$He, $^7$Li and $^9$Be,

b) a smooth composition gradient at the base of the convective zone, and

c) a significant improvement of the sound speed square difference between the seismic Sun and the models in this transition region, when we allow the photospheric heavy element abundance to adjust, within the observational incertitude, due to the action of this mixing process.

The impact on neutrino predictions is also discussed.

\end{abstract}
\vspace{-0.5cm}

%\newpage

\section{INTRODUCTION}

The validation of solar structure is crucial to check the hypotheses 
used in stellar evolution modelling. It allows also  
the helium content at the birth of the Sun to be determined, which is key to galactic 
 enrichment and for predicting the present neutrino emission fluxes. 
 With the support of helioseismology, we are now able to constrain 
 the sophisticated calculations performed to describe the present structure 
 of the Sun. Very recently, a tremendous effort
using ground based networks and space experiments has been made 
 to improve the accuracy of the seismic tool at a level where differences with the  theoretical estimates are 
significant. Consequently, it has been demonstrated that this discipline has the potential to check 
classical stellar structure  for solar-type stars in the hydrogen burning phase (Tomczyk et al. 1995, 
 Kosovishev et al. 1997, Turck-Chi\`eze et al. 1997, Christensen-Dalsgaard 1998...).

In parallel, the development of neutrino detectors in particle physics has 
the objective of determining the properties of these particles. 
In this research, the Sun plays, once again, a specific role as a  source 
of electronic neutrinos $\nu_e$ which could be transformed into other flavors 
 before reaching the detectors on Earth (Hata \& Langacker 1997, Bahcall, Krastev \& Smirnov 1998). The solar 
conditions are really inaccessible in experiments on Earth, so one must predict precisely the emitted neutrino 
fluxes in order to progress in our knowledge of these particles and the solar core.
 
At first glance,  helioseismology shows 
that the standard solar framework is correct 
(Brun, Turck-Chi\`eze \& Morel 1998; hereafter paper I). It is 
nevertheless extremely clear that standard models 
have always failed to answer some well known questions or anomalies
due to its simplified assumptions (Turck-Chi\`eze et al. 1993). These open questions concern 
the effect of the rotation and magnetic field, the history of the angular momentum,
 the real way the convection acts, the unexplained photospheric abundances 
 as the photospheric lithium depletion 
and the presence of  mixing at different locations inside the stars.
Recent results obtained with the SOHO satellite (Fr\"ohlich et al., 1997, 
Kosovishev et al. 1997, Gabriel et al. 1997) and ground networks have demonstrated that there exists 
significant differences 
in the square of the sound speed between the Sun and the models 
(Kosovishev et al. 1997, Turck-Chi\`eze et al. 1997). 
Some of them can be attributed to uncertainties 
in the microscopic description of the Sun, while some others must be 
the sign of missing processes in the standard solar framework. 

The idea of this paper is to introduce a new term in the stellar structure 
equations in order to progress beyond the standard framework. This new term describes
 some physical processes which are presumably present in our star
as suggested by the differential rotation profile.
In this study, we hope to treat 
more correctly the transition layer between the convective and 
 radiative zones, with the help of crucial  measured helioseismic variables. The main 
reasons for our interest in this layer are:
\begin{itemize}
\item The change of rotation rate (solid $\Rightarrow$ differential) 
measured by helioseismology (Thompson et al. 1996, Kosovichev et al. 1997) 
develops probably some turbulent mixing in a shallow layer (tachocline) which 
could be where the magnetic field is anchored and amplified by dynamo process (Sch\"ussler 1987, Choudhuri, 
Sch\"ussler \& Dikpati 1997),
\item The fact that the standard solar model framework including microscopic diffusion and 
element settling has failed to treat this transition correctly 
(paper I). Such a model leads to sharp element 
profiles which create a large bump in the sound speed profile below the convective zone and 
cannot reproduce the observed lithium depletion,
\item The turbulent convective zone surely interacts with the stably stratified radiative 
zone (e.g the Brunt-V\"ais\"al\"a frequency $N^2>0$), developing turbulence, modifying the 
thermal structure of the star (convective penetration) (Zahn 1991), or extending the mixing 
zone (overshooting) (Roxburgh 1997 and Zahn 1998) or generating internal waves (Press 1981, 
Schatzman 1993).
\end{itemize}

Following the description of Spiegel \& Zahn (1992) (hereafter; SZ92) of the motions in this 
so-called tachocline and that of Chaboyer \& Zahn (1992) for the chemical evolution, we 
introduce in the equations of the stellar structure an effective diffusion coefficient 
and build several solar models, which are compared to the helioseismic and surface abundance 
observations.

We first present in section 2 the physical model of the solar tachocline we use and deduce the 
vertical macroscopic diffusion coefficient to be introduced in the time evolution equation of the 
chemical abundances. Then in section 3 we 
present solar models with a mixing occuring in the tachocline, its impact on photospheric abundances, especially for lithium and beryllium and discuss the 
effect of the time evolution of such a mixing from the Sun early phases until 
the present days. In section 4 we comment our results on the changes observed in the sound speed profile, on the role of the solar composition and the recent nuclear reacton rates and on neutrino predictions. Finally we conclude in section 5.

\section{THE SOLAR TACHOCLINE}

The Sun is generally assumed to be in hydrostatic and thermal equilibrium, neglecting the effects of 
rotation and magnetic field. However, models built with these simplifying assumptions do not 
totally agree with the helioseismic data, in particular with those obtained by the satellite SOHO. It appears that macroscopic mixing processes must be taken into account not only in the 
convective zone, but also in the radiative interior. 
Macroscopic mixing may be introduced in the solar models by adding an effective diffusivity $D_T$ 
in the equation governing the time evolution of the chemical species $X_i$. The equation 
becomes:
\begin{equation}\label{eqn: timevol}
\frac{\p X_i}{\p t}=-\frac{\p (4\pi\rho r^2 X_i V_i)}{\p m} +\mbox{nuclear 
terms,}\\
\end{equation} 
where the velocity V$_i$ of species i with respect to the center of mass is:
\begin{equation}
V_i=-4 \pi \rho r^2(D_i+D_T)\frac{\p \ln X_i}{\p m}+v_i.
\end{equation}
The velocity $V_i$ is the sum of a term which depends on the concentration 
gradient and another which does not, $v_i$ (Burgers 1969, Proffit \& Michaud 1991). In this paper, 
we add a macroscopic diffusion term $D_{T}$ to the microscopic diffusion term $D_{i}$. This will 
describe the mixing occurring in the shear layer connecting the differential rotation in the 
convection zone to the solid rotation in the radiative interior.
As presented in SZ92, a possible physical interpretation of this tachocline is obtained by 
invoking anisotropic turbulence with much stronger viscous transport in the horizontal than in 
the vertical direction (i.e $\nu_H>>\nu_V$). Such turbulence reduces the differential rotation 
and therefore inhibits the spread of the layer deep inside the radiative zone, leading to a stationary state. 
For a justification of this anisotropy, we refer to Michaud \& Zahn (1998).

\subsection{The hydrodynamical description}\label{section_tac}

 The conservation of mass, momentum and entropy is given by:
\begin{equation}
\frac{\partial \rho}{\partial t}+\vec{\nabla}.(\rho \vec{V}) = 0 
\end{equation}
\vspace{-0.5cm}
\begin{eqnarray}
\rho \left(\frac{\partial \vec{V}}{\partial 
t}+(\vec{V}.\vec{\nabla})\vec{V}+2\vec{\Omega}\wedge\vec{V}+ 
\vec{\dot{\Omega}}\wedge\vec{r}\right) 
%\nonumber \\
 = -\vec{\nabla}P \nonumber \\
-\rho\vec{\nabla}\Phi+\vec{\nabla}.\parallel\tau\parallel 
\end{eqnarray}
\begin{equation}
\rho T \left(\frac{\partial}{\partial t}+\vec{V}.\vec{\nabla} \right)S = 
\vec{\nabla}.(\chi \vec{\nabla}T)
\end{equation}
where $\rho$ is the density, $P$ the pressure, $T$ the temperature, 
$\vec{V}=(u,v,r\hat{\Omega}\sin\theta)$ is the local velocity in the rotating 
frame $\vec{\Omega}$. In this expression, the differential rotation $\hat{\Omega}$ appears explicitly,
 $S$ is the specific entropy, 
$\parallel\tau\parallel$ the viscous stress tensor and $\Phi$ the gravitational 
potential. In order to ease the solution of this system of equations, some simplifying assumptions were made 
by 
SZ92, mainly:

\begin{itemize}
\item velocities are small compared to the sound speed: $\p \rho/\p t$ is negligible (anelastic 
approximation), 
\item advection terms and viscous forces are small compared to the Coriolis acceleration (i.e 
small Rossby and Ekman numbers),
\item the tachocline is thin compared to the pressure scale height and the turbulent horizontal 
viscosity is stronger than the vertical one (i.e $\nu_H>>\nu_V$).
\end{itemize} 

After separating each variable into its mean value on the sphere plus a 
perturbation, e.g. $T(r,t)+\hat{T}(r,\theta,t)$, the linearized form of 
the equations in a stationary state is given by SZ92. Spiegel and Zahn define a streamfunction 
$\Psi$ for the meridional flows by:
\begin{equation}
r^2 \rho u=\frac{\p \Psi}{\p x} \mbox{ , }r \rho v \sin \theta =\frac{\p \Psi}{\p r}
\end{equation}

\noindent
where $x=\cos\theta$, and project  the variables of the equations describing the tachocline on  
horizontal eigenfunctions $F_i$:
\begin{eqnarray}\label{eqn: fi}
(\hat{P},\hat{T},u)&=&\sum_i (\tilde{P}_i,\tilde{T}_i,u_i)F_i(x) \nonumber \\
\Psi&=&\sum_i \tilde{\Psi}_i \int F_i(x)dx \\
x\hat{\Omega}&=&\sum_i \tilde{\Omega}_i \frac{d F_i}{dx}. \nonumber
\end{eqnarray}
The dominant term for the even eigenfunctions is $i=4$, but we will consider in our calculation 
even terms up to $i=8$. From this projection, SZ92 deduce a fourth-order differential equation 
for the modal function $\tilde{\Omega}_i(r)$, defined in eq (\ref{eqn: fi}):
\begin{eqnarray}
\frac{4\Omega^2}{\mu_i^4}\frac{1}{\rho \nu_H}\frac{\p}{\p 
r}\left\{\frac{g}{N^2c_pT}\frac{\p}{\p r}\left[\chi r^2 \frac{\p}{\p r} \left( 
\frac{T}{\rho g} \frac{\p}{\p r} \rho r^2 \tilde{\Omega}_i \right) \right] 
\right\} \nonumber \\ +\tilde{\Omega}_i=0.
\end{eqnarray}
where $\mu_i$ are eigenvalues related to the eigenfunctions $F_i$ (for $i=4$, $\mu_4=4.933$), 
$\chi=16\sigma T^3/3\kappa\rho$ is the radiative conductivity, $N$ the Brunt-V\"ais\"al\"a frequency,
$\Omega$ the mean rotation rate of the radiation zone, 
and all the other variables have their usual meaning.

The problem is simplified by treating the tachocline as a boundary layer in 
which the rapidly varying quantity is $\tilde{\Omega}_i$ and all the other 
physical quantities are assumed constant. By doing so, an analytical solution of the simplified 
fourth-order differential equation is (for simplicity we only give the expression for the modal 
index $i=4$):
\begin{equation}\label{rot}
\tilde{\Omega}_4(\zeta)=\tilde{\Omega}_4(0)\sqrt{2} \exp(-\zeta)\cos(\zeta-\pi/4)
\end{equation}  
where $\tilde{\Omega}_i(0)=Q_i\Omega$, $Q_i$ are  numerical coefficients depending on the latitudinal 
variation of the differential rotation and we have defined a non dimensional depth:
\begin{equation}\label{zeta}
\zeta=\mu_i (r_{bcz}-r)/d
\end{equation}
with,
\begin{equation}\label{eqn: dd}
d=r_{bcz}(2\Omega/N)^{1/2}(4K/\nu_H)^{1/4}
\end{equation}
being a parameter related to the tachocline thickness $h$, $r_{bcz}$ the radius   
and $K=\chi/\rho c_p$ the radiative diffusivity, at the base of the convective zone. 

With the latitudinal dependence of the angular velocity at the base of the convection zone 
borrowed from Thompson et al. (1996): 
$\Omega_{bcz}/2\pi=456-72x^2-42x^4$ nHz, we have reestimated the  coefficient 
$Q_4=-1.707\times10^{-2}$, as well as the ratio between the rotation rate of the radiative zone and 
the equatorial rate: $\Omega/\Omega_0=0.9104$. The prediction by Gough \& Sekii (1997), who 
consider instead the magnetic stresses, is $\sim 0.96$; presently, the seismic observations 
suggest a rotational ratio of $0.94\pm0.01$ (Corbard et al. 1999), which is intermediate 
between these two theoretical estimates.

The inversion of the rotational profile from helioseismology observations
 gives the width of the tachocline and its location below the convection zone. For a comparison 
between the observational thickness and the theoretical one, we have to consider the first zero 
of equation (\ref{rot}), e.g $h=3 \pi d/4 \mu_4\sim d/2$. The parameter $d$ is then 
approximately equal to $2h$. In the solar models of section \ref{section_sol}, we will treat 
$h$ (hence $d$) as an adjustable parameter, in order to agree with the helioseismic inversion 
of the tachocline thickness (Basu 1997, Charbonneau et al. 1998, Corbard et al. 1998, 1999).

 Our purpose now is to obtain an analytical expression for the vertical velocity $u$. We 
already have an expression for the differential rotation, the next step is to connect $u$ to 
$\hat{\Omega}$. From the conservation of the angular momentum, we find a relation between the 
stream function $\hat{\Psi}$ and the differential rotation $\hat{\Omega}$. By writing the streamfunction as 
$\tilde{\Psi}_i=\rho r^2 u_i$ and introducing the analytical expression of $\tilde{\Omega}_i$ (eq \ref{rot}), 
we finally obtain after integration, the radial dependence of $u_i$; in the case of $i=4$ we get:
\begin{equation}\label{u4}
u_4(\zeta)=\frac{1}{2} \frac{\nu_H d}{r_{bcz}^2} \mu_4^3 Q_4 \exp(-\zeta) cos(\zeta) .
\end{equation}

 From this equation (\ref{u4}), we shall derive an expression for the macroscopic diffusivity $D_T$ of the 
motions occurring in the tachocline.

\subsection{Macroscopic diffusion of chemical species}
 We determine in this section 
the influence of the mixing on the evolution of chemical species, 
using the expression for the vertical velocity amplitude $u_i$ (eq. \ref{u4}). The 
anisotropic diffusion, invoked by SZ92 to stop the spread of the tachocline deeper in the 
radiative zone, interfers with the advective transport of chemicals. Chaboyer \& Zahn (1992) 
have shown that the result is a diffusive transport in the vertical direction, with an 
effective diffusivity given by:
\begin{equation}
D_{T}=\frac{r^2}{D_H}\sum_n \frac{U^2_n(r)}{n(n+1)(2n+1)} 
\end{equation}
where $U_n$ are the coefficients of the expansion of the vertical component of 
the velocity $u$ in Legendre polynomials.
The eigenfunctions $F_i$ introduced in section \ref{section_tac} may be 
projected on these Legendre polynomials, and to a good approximation one has 
(keeping only the n=4, 6 and 8 terms)
\begin{equation}\label{precoef}
\hspace{-0.5cm}D_T=\frac{r^2}{D_H}\left[\left(\frac{8}{3}\right)^2 
\frac{u^2_4(r)}{180}+\left(\frac{16}{5}\right)^2 
\frac{u^2_6(r)}{546}+\left(\frac{128}{35}\right)^2 \frac{u^2_8(r)}{1224}\right]
\end{equation} 

\begin{figure*}[!ht]
\setlength{\unitlength}{1.0cm}
\begin{picture}(8,7.5)
\includegraphics{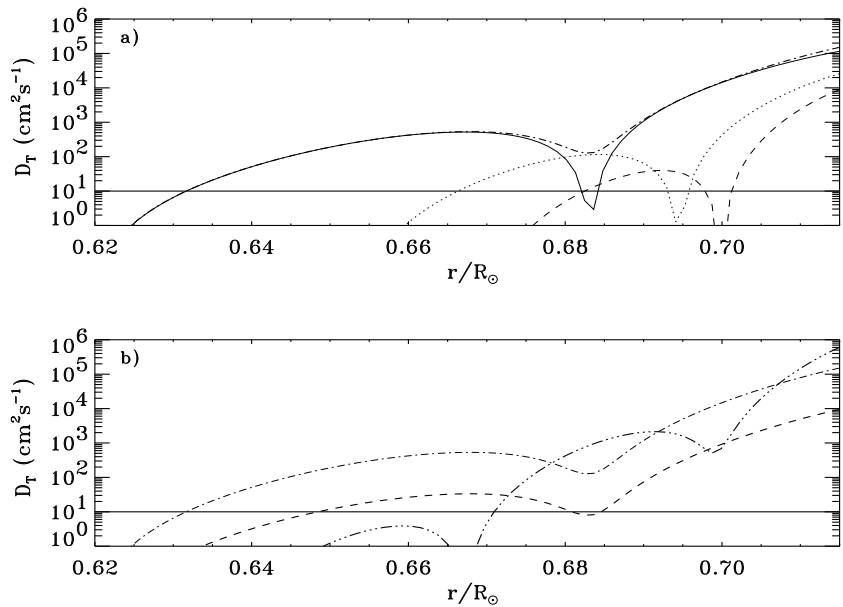}
\end{picture}      
\caption{\label{f0} Effective diffusivity $D_T$ for tachocline mixing: a) Model $B$ ($N=25$ and $d=0.1$) for 
index $n=4$ $B_4$ (solid line), $n=6$ $B_6$ (...), $n=8$ $B_8$ (- -) and 
$\sum_{n=4}^{8} B_n$ (thick -.-). b) Models $A$ ($N=100$ and $d=0.1$) (- -), 
$B$ ($N=25$ and $d=0.1$) (-.-) and $C$ ($N=25$ and $d=0.5$) (-...-). The horizontal solid line represents approximately the microscopic diffusion $D_i$.}
\end{figure*}

One reaches the following expression for the vertical diffusivity in replacing $u_4$ by 
equation (\ref{u4}) in equation (\ref{precoef}), same for $n=6$ and $n=8$ 
(c.f Fig \ref{f0}):
\begin{eqnarray}\label{eqn: coef}
D_T(\zeta)=\frac{1}{180}\frac{1}{4}\left(\frac{8}{3}\right)^2 \nu_H 
\left(\frac{d}{r_{bcz}}\right)^2 \mu_4^6 \, Q_4^2 \nonumber \\ \times \exp(-2\zeta) \cos^2(\zeta) + \mbox{ higher order terms.}
\end{eqnarray}
where $\zeta$ is the scaled vertical coordinate defined in equation (\ref{zeta}).

In this expression,  it is assumed that the horizontal component of the macroscopic diffusivity $D_H$ is equal to 
the horizontal viscosity $\nu_H$. This coefficient is then introduced in the diffusive part of the equation 
(\ref{eqn: timevol}) for the time evolution of the chemical abundances $X_i$,  
and we could thus obtain solar models including a treatment of the tachocline.
Fig \ref{f0} compares the different values of the macroscopic diffusivity for models discussed in the next 
section, to the microscopic diffusion (here represented as a horizontal solid  line). The value of 10 $cm^2 s^{-1}$ for $D_i$ is deduced from the expression of Michaud \& Proffitt (1993) for $^4$He, in the case of the reference solar model at the present age. It must be pointed out that in our calculation these microscopic diffusion coefficients as well as the microscopic velocity $v_i$ are evaluated for all the chemical species present in our gaz mixture, and are not assumed constant (see paper I). The amplitude of $D_{^4He}$ varies only by 10\% in the range [0.6,0.72] $R_{\odot}$ and doesn't change significantly with time. The oscillating behaviour of the macroscopic coefficient seen in Figure (\ref{f0}) reflects the sub cells structure of the tachocline and reduces the lithium depletion in comparison with a purely exponential coefficient by a factor $\sim$ 1.5-2 and induces stages in the radial chemical composition profile (c.f 3.2 and Fig \ref{f3}). The mixing in the tachocline thus depends on the following parameters, for a given model of the present Sun:
\begin{itemize}
\item the rotation rate $\Omega$, a measured quantity;
\item the differential rotation rate $\hat{\Omega}$ (or its projection $Q_i$) inferred from helioseismology;
\item the Brunt-V\"ais\"al\"a frequency $N$, determined by the extent of convective penetration;
\item the horizontal diffusivity $\nu_H$, which presumably sides with the differential rotation rate.
\end{itemize}
There are at present no firm prescriptions available for $N$ and $\nu_H$, therefore we must consider these or two combinations of them, as free parameters in our calculations. We have chosen to keep $N$ and to take $d$ (twice the thickness of the tachocline), because both of them are constrained by upper limits drawn from helioseismology.

\section{SOLAR MODELS}\label{section_sol}

\subsection{Description of the models}

In paper I (Brun, Turck-Chi\`eze \& Morel 1998), we have calculated a new standard model using 
the CESAM code (Morel 1997) and have discussed the validity of such a model in the light of the 
present helioseismic observations. We just recall here the main physical inputs present in this 
reference model. We used the solar chemical composition of Grevesse, Noels \& Sauval (1996), 
the OPAL96 opacity tables  Iglesias \& Rogers (1996) and  the equation of state of Rogers, 
Swenson \& Iglesias (1996),  the recent updated nuclear reaction rates of 
Adelberger et al. (1998), the prescription of Mitler (1977) for the treatment of the nuclear 
screening reaction rates for a charge Z$>$2 (Dzitko et al., 1995) and the microscopic diffusion 
coefficients suggested by Michaud \& Proffitt (1993). 

 In the present paper (called paper II), we use the ``pure microscopic diffusion''
  model as the  reference model. We then compute  solar models using the 
  macroscopic coefficient $D_T$ described by equation (\ref{eqn: coef}) and 
 follow the time evolution of these solar models from the pre-main sequence 
  (PMS) until an age of 4.55 Gyr, making the age of our present Sun 4.6 Gyr 
  (including about 0.05 Gyr of pre-main sequence). The adaptable time step 
  increases from $\sim$20 years up to 200 Myr during the evolution. We start 
  from the PMS so as to properly follow the $^7$Li abundance evolution, including the 
  macroscopic diffusion coefficient along all the solar phases. We 
  introduced an updated nuclear reaction rate for $^7$Li(p,$\alpha$)$^4$He, 
  Engstler et al. (1992) instead of the Caughlan \& Fowler (1988) one. The 
  results of the solar models we calculated for this study are summarized in 
  Tables \ref{table 0} and \ref{table 1} and  the photospheric abundances of the elements which 
  are sensitive to the
  adding macroscopic process are compared to observations in Table \ref{table 2}.

\subsection{Time independent rotation}\label{sect: notime}

We first compute solar models including the effective diffusivity $D_T$ (eq. \ref{eqn: coef}), 
with a tachocline thickness $h$ of 0.05 $R_{\odot}$ ($d=0.1$ 
$R_{\odot}$) in agreement with the helioseismic measurements (Basu 1997, Corbard et al. 1998, 1999, 
Charbonneau et al. 1998), a Brunt-V\"aiss\"al\"a frequency $N$ of 100 or 25 $\mu$Hz to take into account the rapid variation of this quantity at the base of the convective zone, and 
a rotation $\Omega$ at the base of the convective zone of 0.415 
$\mu$Hz; we call these coefficients $A$ and $B$ respectively (see 
Table \ref{table 0}). The reference model has a present photospheric 
helium abundance of 0.2427 in mass, corresponding to an $^4$He diffusion of 
10.8\%. This value of Y$_s$ is slightly too low if we compare with the Basu 
\& Antia (1995) value for the OPAL equation of state, Y$_s=0.249 \pm 0.003$. 
When introducing the macroscopic diffusion coefficient, we mix back helium in the convection
 zone, inhibiting the microscopic diffusion by 15\% or 25\% and leading to a 
 present photospheric $^4$He abundance Y$_s=0.2452$ or Y$_s=0.2473$ 
 respectively for models $A$ and $B$ (c.f Tables \ref{table 0} and 
 \ref{table 2}, see also Brun, Turck-Chi\`eze \& Zahn 1998). This is 
 in better agreement with the helioseismic values. The decrease of the 
 Brunt-V\"ais\"al\"a frequency $N$ leads to an increase of the efficiency 
 of the mixing in the tachocline layer and reduces the settling of the chemical 
 species. We also note that the initial helium content Y$_0$ is smaller than 
 0.27 in mass (cf Table \ref{table 0} and also paper I).

\begin{figure*}[htb*]
\setlength{\unitlength}{1.0cm}
\begin{picture}(9,7)
\includegraphics{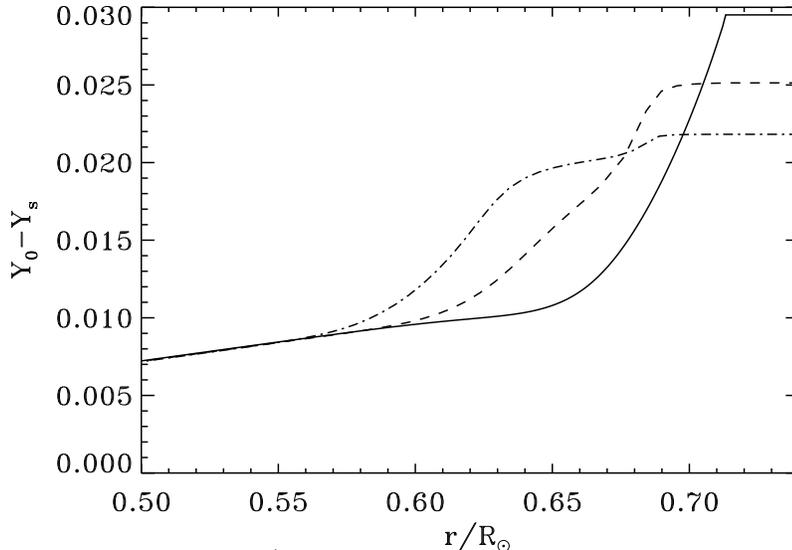}
\end{picture}      
 \caption{\label{f3} Radial profile of the difference of  $^4$He composition 
 between the initial and present values for the reference solar model including
  only microscopic diffusion (solid line) and solar models where we add a 
  macroscopic mixing due to the presence of the tachocline: coefficient $A$ 
  (- -) and $B$ (-.-) (see section \ref{sect: time} and Tables \ref{table 0}, \ref{table 2} for the model characteristics).}
\end{figure*}

 In Figure \ref{f3} we present the difference between the initial helium composition Y$_0$ 
and the present photospheric one Y$_s$ for several models 
including the reference one. Before the introduction of microscopic diffusion, this 
difference was null because the present photospheric values were adopted as the initial 
composition of the star. However now, when we compute a standard model including the microscopic 
diffusion and the settling of chemical species, the initial abundances are assumed to be different from their 
present day values and there is an iteration on the composition to get the photospheric Z/X 
ratio in agreement with the one measured at the present time: Z/X= 0.0245 $\pm$ 0.003.

The prominent feature of Figure \ref{f3} is the smoother composition 
profile for models including the macroscopic diffusivity $D_T$, in comparison 
with the reference model, which presents a sharp composition gradient just 
below the convective zone. The composition profile is also almost constant 
below the convective zone over a distance which is related with the value 
of $d$ and $N$. A smaller Brunt-V\"aiss\"al\"a frequency value induces a greater mixing, 
and then an extended plateau. The characteristic steps in the radial profile come from the 
cosine term in the expression of $D_T$ (eq. \ref{eqn: coef}) (see also Elliott (1997)). An 
explanation could be that the simplified assumption in the treatment of the tachocline of SZ92, 
of a constant turbulent horizontal viscosity $\nu_H$, induces this radial dependence of the 
coefficient $D_T$ and thus the steps in the composition profile. A more careful study should be 
done to verify this statement.\\
 Concerning the $^7$Li and $^9$Be depletion, we see in Table 
 \ref{table 2} that the mixing increases the lithium depletion by a factor up 
 to $\sim 4$, without burning $^9$Be, which agrees with the revised determination by Balachandran \& Bell 
 (1998). Nevertheless, such an increase of the $^7$Li depletion is still 
 insufficient to reproduce the photospheric lithium depletion (Grevesse, Noels 
 \& Sauval 1996, Cayrel 1998 and references therein). 
%\vspace{0.5cm}
\subsection{Time dependent rotation}\label{sect: time}

In this section we shall take into account the time evolution of the tachocline width 
and the efficiency of the mixing from the early phase of the Sun until now. 
In the previous section, we have used a constant value for $d$ based on 
the present observations of the tachocline thickness  in the effective diffusion coefficient $D_T$. 
Thus, in order to improve the description of tachocline mixing, 
we now introduce  a time dependence of the coefficient $D_T$ and of the tachocline thickness $h$ (hence $d$), 
related to the global and the differential rotation of the Sun as follows:
\begin{equation}
\hspace{-0.5cm}D_T \propto \nu_H \left(\frac{d}{r_{bcz}}\right)^2 Q_i^2 \propto \Omega \nu_H^{1/2} (\hat{\Omega}/\Omega)^2 \mbox{ , \ } d \propto 
\Omega^{1/2}/\nu_H^{1/4}, \nonumber
\end{equation}
where we have used equations (\ref{eqn: coef}) and (\ref{eqn: dd}). 
Assuming that the turbulent viscosity is proportional to the differential rotation (i.e 
$\nu_H\propto\hat{\Omega}$) and introducing the dependence of differential rotation on
rotation observed by Donahue, Saar \& Baliunas (1996) 
($\hat{\Omega}\propto\Omega^{0.7\pm0.1}$), we finally obtain for $D_T(\Omega)$ and $d(\Omega)$ the following 
scaling laws:
\begin{equation}\label{eqn: omegat}
D_T \propto \Omega^{0.75\pm0.25} \mbox{ , \ \ \ } d \propto \Omega^{(1.3\mp0.1)/4}. 
\nonumber
\end{equation}
We conclude that the tachocline mixing was stronger in the past both because that layer was thicker and because 
the diffusivity was larger. Note that the observational uncertainties in the relation between $\hat{\Omega}$ and $\Omega$ act in opposite way on the dependence of  
$D_T(\Omega)$ and $d(\Omega)$ (i.e a stronger mixing implies a smaller tachocline extent). We render the mixing in the tachocline time-dependent, through 
$D_T(\Omega(t))$ and $d(\Omega(t))$, by using the spin-down law $\Omega(t) \propto t^{-1/2}$ which was 
deduced by Skumanich (1972) from the rotation rate of stellar clusters of different ages.

Models $A_t$, $B_t$ and $C_t$  correspond to solar evolutions
 with time-dependent coefficients given by equation 
(\ref{eqn: omegat}) and different values for $N$ and the present tachocline thickness $d$ (see Table 1 and 
2). The first two cases 
correspond to the same values of $d$ and $N$ that existed in the previous section but with a time 
dependence of the coefficient $D_T$. The third shows the influence of  a thinner tachocline ($d=0.05 
R_{\odot}$) as suggested by more recent studies (Corbard et al. 1999). The introduction of this 
time-dependence increases the 
inhibition of the microscopic diffusion of helium during the evolution in 
comparison with the previous solar models considered in section \ref{sect: 
notime} (see Table \ref{table 2}). These models show an inhibition of $^4$He 
up to 27\% for model $B_t$ ($d=0.1$, $N=25$) leading to a surface helium 
Y$_s=0.2477$, 
very close to the seismic value of Basu \& Antia (1995) . 

\begin{figure*}[!htb]
\setlength{\unitlength}{1.0cm}
\begin{picture}(8,7)
\includegraphics{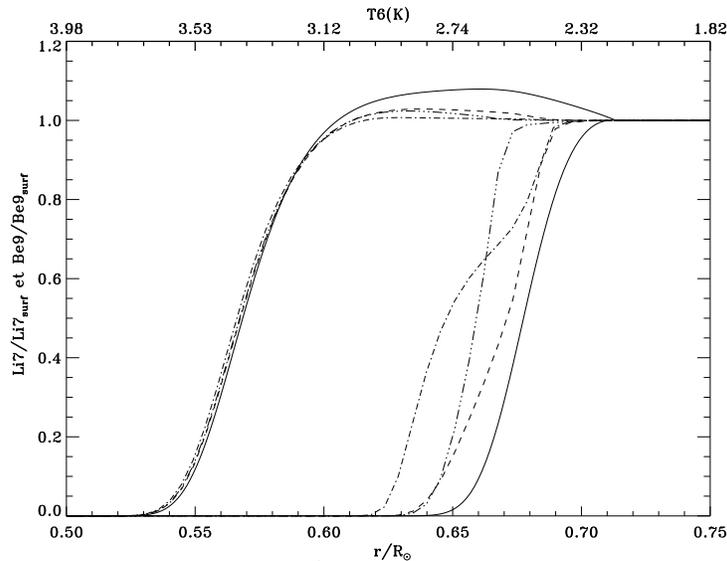}
\end{picture}
\vspace{0.2cm}
 \caption{\label{f5}   $^7$Li and $^9$Be radial profile  for several models (superimposed the 
corresponding temperature T$_6=$T/$10^6$ K obtained for the reference model). Respectively: 
reference (solid line), models $A_t$ (dash), $B_t$ (dash dot) and $C_t$ (dash three dots). The mixing below the convective zone 
modifies only the $^7$Li profile and not the $^9$Be one (except the flat 
plateau for the mixed models).}
\end{figure*}

We plot the $^7$Li and $^9$Be radial profiles in Figure \ref{f5},  
for the reference model and the time dependent models 
calibrated in Z/X  at 4.6 Gyr, with the corresponding radial temperature scale. These two light chemical 
elements are sensitive to mixing processes 
occuring in the stars because their nuclear burning temperatures are very low 
(resp. 2.5 $10^6$ and 3.2 $10^6$ K) (Baglin \& Lebreton 1990). It is well 
known that the solar $^7$Li is depleted by a factor $\sim$ 100 (Grevesse, Noels \& Sauval 1996)  and recent 
revised abundance determinations  have shown  that  $^9$Be abundance is practically not depleted 
(Balachandran \& Bell 1998) contrary to what was thought previously. 
These new observational constraints can only be satisfied if those chemical
 species are mixed in a rather thin layer below the convective zone, which 
 is the case for the tachocline mixing considered here.

We clearly see in Figure \ref{f5} that this mixing process modifies the 
distribution of lithium but not of beryllium (except the flat
 plateau for the mixed models in comparison with the ``pure'' diffusion one). 
 These results are 
confirmed in Table \ref{table 2} where we show the initial over present
 ratio of the $^7$Li and $^9$Be. We also remark that models including a 
macroscopic diffusivity with a low value of the Brunt-V\"ais\"al\"a frequency 
$N=25$ (i.e $B_t$ and $C_t$) burn more $^7$Li, 
because of their higher absolute value ($\geq 10^5$ cm$^{2}$s$^{-1}$) 
in comparison with the microscopic diffusion one  ($\sim 10$ cm$^{2}$s$^{-1}$),
 than models with a higher $N=100$ (i.e $A_t$) (Figure 1). The models $B_t$ and 
 $C_t$ strongly inhibit the settling of the chemical species and mix all 
 the elements over a distance where the lithium burning temperatures are 
 reached $r \sim 0.67 R_{\odot}$. The characteristic time for this mixing,
  for a $D_{T}$ of $10^5$ cm$^{2}$s$^{-1}$ is $\sim 3$ Myr. This time is 
  intermediate between the convective cell turnover time ($\sim 1$ month) and
   the microscopic diffusion one ($> 10$ Gyr). We also see in Table \ref{table 2} 
   that the time dependent diffusivity improves the $^7$Li depletion, where a value of 
   $\sim$ 100 is obtained without destroying $^9$Be and without increasing the $^3$He/$^4$He surface 
ratio over the past 3 Gyr, as deduced by Geiss \& Gloeckler (1998) from meteorites and solar 
wind abundance measurements. As already mentioned these abundance constraints can only be satisfied if the 
mixing occurs in a thin layer below the convective zone, which is the case for the tachocline considered 
here. 

\begin{figure*}[!htb]
\setlength{\unitlength}{1.0cm}
\begin{picture}(8,6.6)
\includegraphics{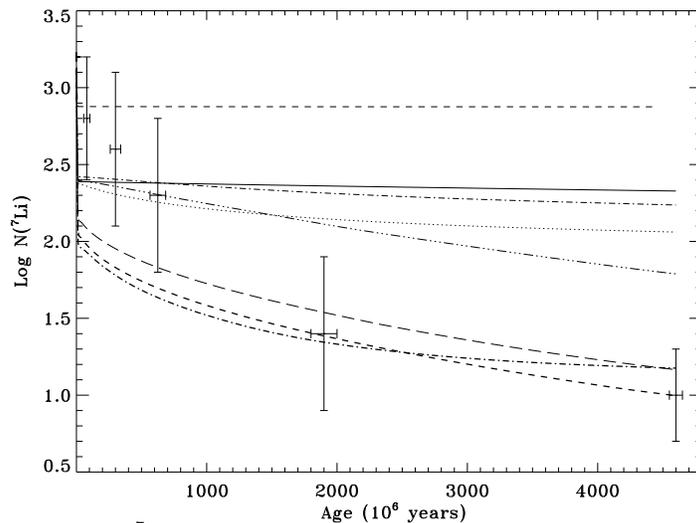}
\end{picture}
%\vspace{0.2cm}
 \caption[]{\label{f6} Time dependent depletion of $^7$Li for several solar 
models: no microscopic diffusion (- - -), with microscopic diffusion (solid line) and with 
mixing in the tachocline thickness: models $A$ (-.-), $B$ (-...-), then with 
time dependent mixing, models $A_t$ (...) and $B_t$ (long dash), $C_t$ (thick -.-) and 
$B_{tz}$ (Z$_0=$Z$_0^{std}=0.01959$) (thick - -). We superimposed on the theoretical curves 
the open cluster observations: ($\alpha Per$: Balachandran et al. (1996); Pleiades: Soderblom 
et al. (1993b); UMaG: Soderblom et al. (1993a); Hyades: Thorburn et al. (1993); NGC752: Balachandran (1995).  (Figure adapted from Vauclair \& Richard 1998 and cluster age uncertainties deduced from Lebreton et al. 1997 and private communication.) }
\end{figure*}

In Figure \ref{f6} we present the lithium depletion occurring during the Sun 
evolution for the different models presented, plus one without microscopic diffusion. Clearly, only the 
diffusion models including mixing in the tachocline show a substantial depletion during the main sequence evolution as it is observed in different clusters. One 
notes that the lithium depletion is largely dependent on the Brunt V\"aiss\"al\"a frequency 
adopted at the base of the convective zone and on the time dependence of the macroscopic coefficient. The 
modification of  the tachocline thickness has only little impact on the lithium 
depletion as the higher value of the coefficient largely compensates the reduction of the 
thickness. However one sees that the $^7$Li depletion slopes in these two models (i.e $B_t$ and 
$C_t$) are not the same. The strong time dependent mixing with $N=25$ presents a correct value of the 
solar 
$^7$Li depletion. The lithium depletion during the PMS is probably overestimated due to the crude spin-down 
law we have adopted. A more 
detailed analysis of these phases is under study (see also Vauclair \& Richard 1998 and Montalban \& 
Schatzman 1996). 

\section{DISCUSSION}

\subsection{ The sound speed profile}

\begin{figure*}[!ht]
\setlength{\unitlength}{1.0cm}
\begin{picture}(8,7.5)
\includegraphics{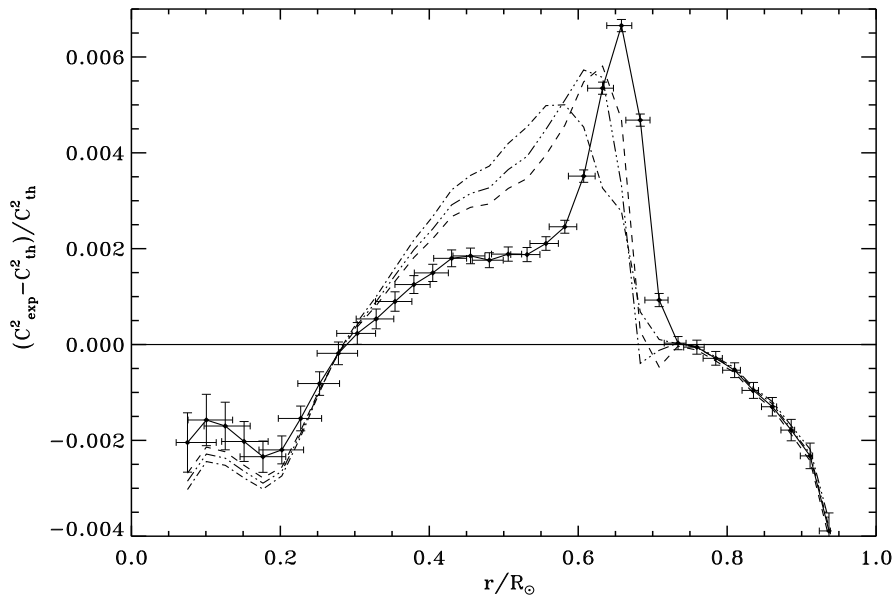}
\end{picture}      
\caption{\label{f2} Sound speed square difference between GOLF+MDI data, the reference model 
(solid line) and models including a macroscopic term: $A_t$ (- - -), $B_t$ (-.-), $C_t$ (-...-) (see parameters in Table 2).}
\end{figure*}

We now compare the sound speed profile of the different models with the helioseismic inversion 
results obtained with the GOLF+MDI data (Lazrek et al. 1997, Roca Cortes et al. 1998, 
Kosovishev et al. 1997) and described in Turck-Chi\`eze et al. (1997, 1998) (Fig 
\ref{f2}), we see that the peak localized at $r \sim 0.68 R_{\odot}$ is practically erased by 
this process, depending on the strength of the mixing. 

\begin{figure*}[htb*]
\setlength{\unitlength}{1.0cm}
\begin{picture}(8,7.5)
%\begin{picture}(12,6.5)
\includegraphics{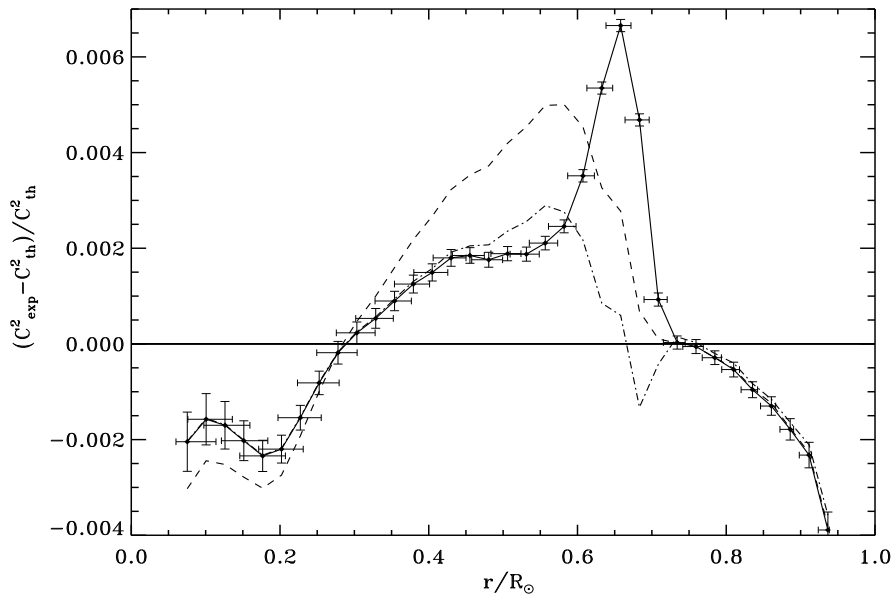}
\end{picture}
\caption{Sound square speed difference between GOLF+MDI data, the reference model (solid line) 
and two models including a macroscopic term: $B_t$ calibrated in Z/X (- -) and  $B_{tz}$ with a non calibrated 
Z$_0=$Z$_0^{std}=0.01959$ (-.-). \label{f1}}      
\end{figure*}

\begin{figure*}[!ht]
\setlength{\unitlength}{1.0cm}
\begin{picture}(8,13)
%\special{psfile=fig7a.ps hoffset=410 voffset=-40 hscale=45 vscale=43 angle=90}
%\special{psfile=fig7b.ps hoffset=406.5 voffset=-250 hscale=45 vscale=43 angle=90}
\includegraphics{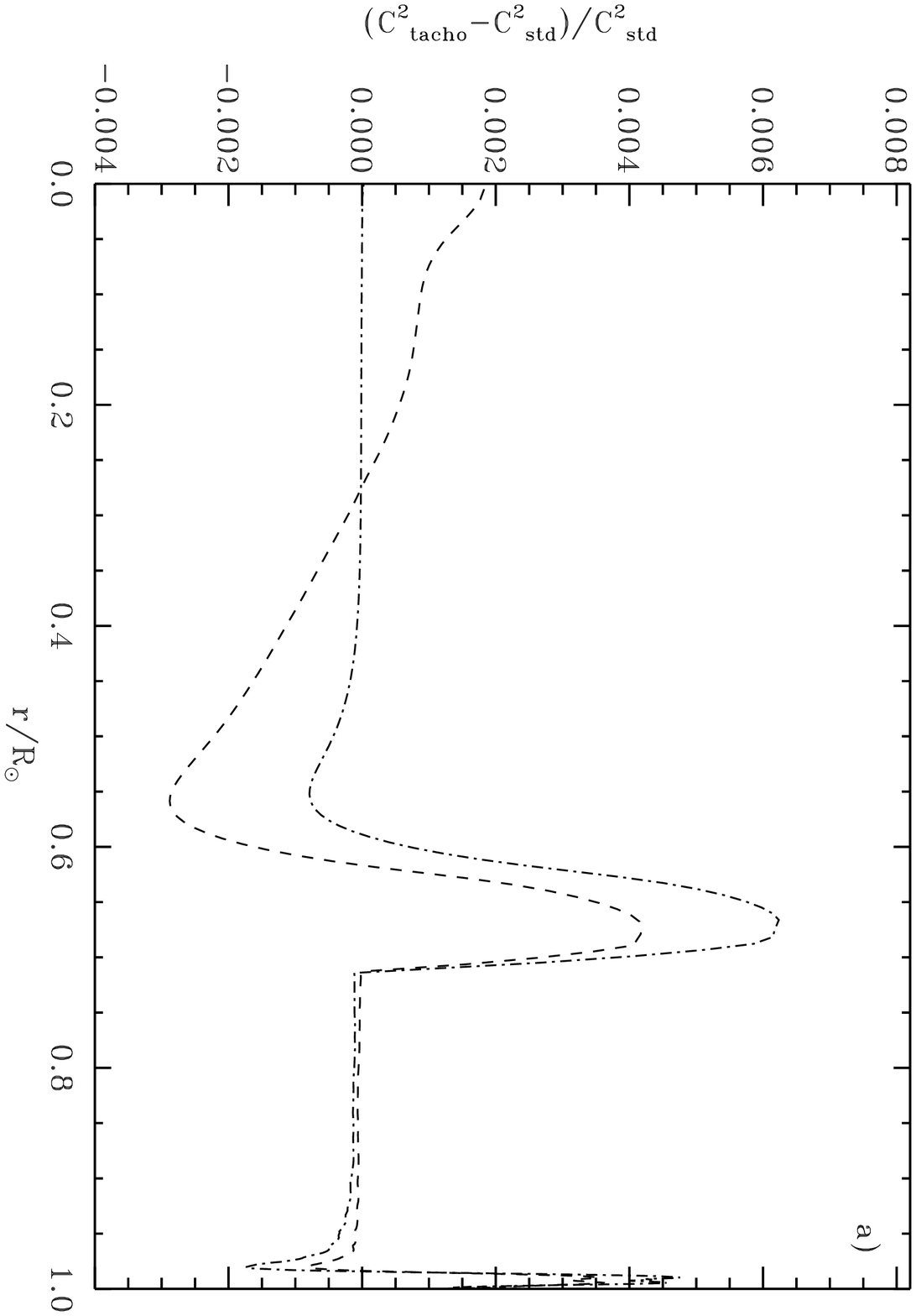}
\includegraphics{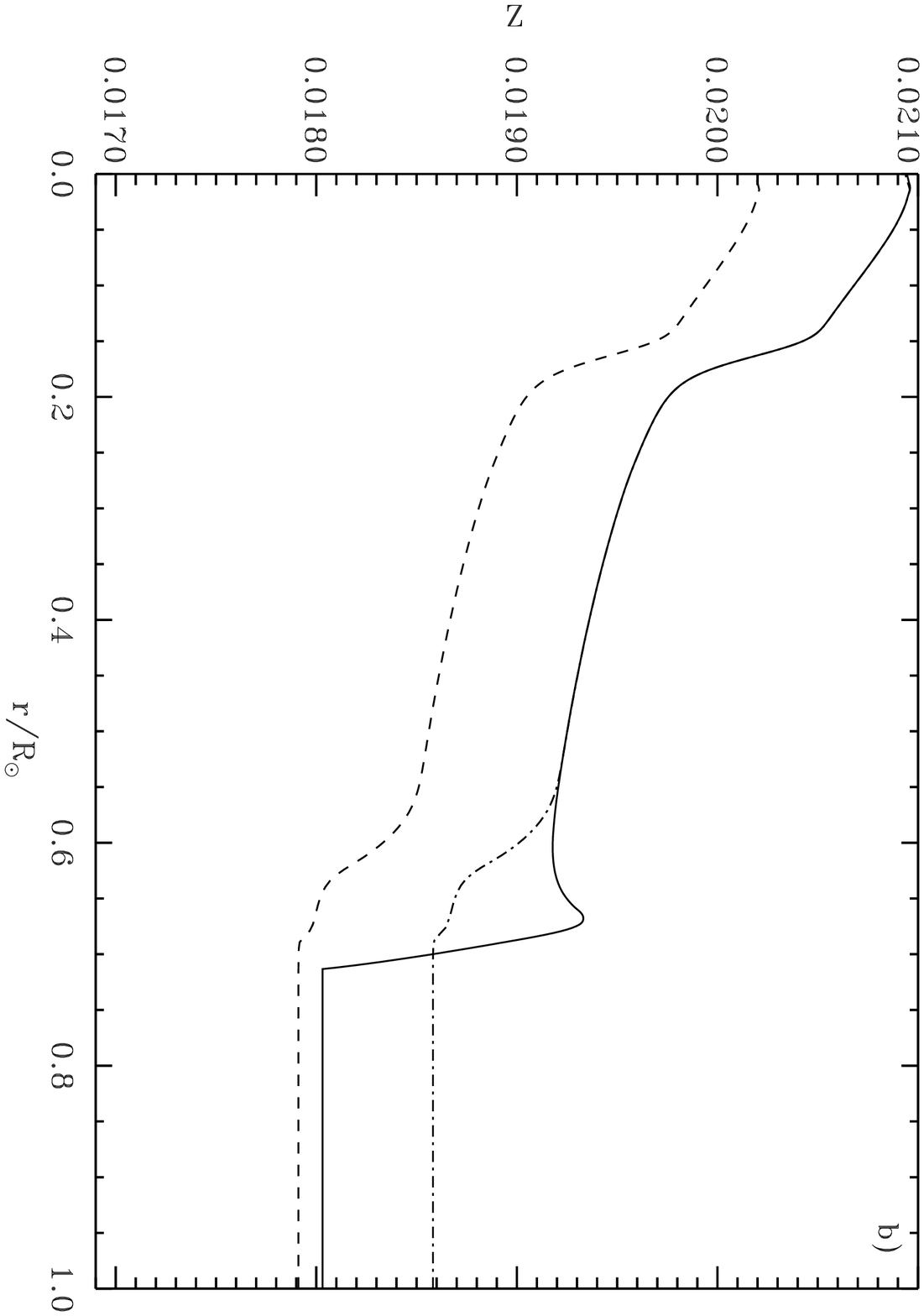}
\end{picture}
\caption{{\em a)} The effect of calibration on the sound square speed difference 
between the reference model and two models including a macroscopic term: $B_t$ calibrated in Z/X (- - -) and $B_{tz}$
 with a non calibrated Z$_0=$Z$_0^{std}=0.01959$ (dash dot line). {\em b)} Heavy Element Composition for the reference model including only 
microscopic diffusion (solid line) and two models with macroscopic mixing due to the tachocline: $B_t$ calibrated in Z/X (- -) and $B_{tz}$ with a non calibrated Z$_0=$Z$_0^{std}=0.01959$ (-.-). 
(see Table \ref{table 1} for the model characteristics). \label{f4}}
\end{figure*}

In all the models shown in this figure (\ref{f2}), we have adjusted the initial composition in order to get at the solar age, 
 the present observed photospheric heavy element/hydrogen ratio, i.e  
 Z/X=0.0245. This 
calibration causes the swell  of the small bump located  
between the nuclear core and the convective zone, since it reduces the initial metallicity (see Table 2). 
It is well known that the heavy element abundance has a direct influence on the solar structure,
mainly through the opacity coefficients and hence it influences the sound speed profile between 0.4 and 0.7 $R_{\odot}$
(Turck-Chi\`eze 1995, 1998 and paper I). The result is quite different if we do not constrain the photospheric Z/X value of the mixed model, starting with the 
heavy element composition Z$_0=0.01959$ of the reference model for example, we obtain a different chemical 
composition at the solar age (see for example columns 3 and 4 of Table \ref{table 1} for the 
coefficients $B_t$ and $B_{tz}$).
In Figure \ref{f1}, we draw the corresponding squared radial sound speed  differences between 
the models and the induced solar value for two models where we have kept (i.e coefficient $B_{t}$) or relax ($B_{tz}$) this surface 
heavy element constraint. We see that in both cases where the macroscopic term is introduced, the 
bump around 0.68 $R_{\odot}$ is practically erased by the introduction of the tachocline mixing, 
but the calibrated model is modified along the whole structure and not only close to the bump.  
It is not the case for the non-calibrated model where Z$_0$ is fixed. This is partly 
illustrated by Figure \ref{f4}a, where the non calibrated model changes only 
locally the 
structure of the solar model (dash dot line).

A similar result is obtained by Elliott, Gough \& Sekii (1998) invoking an advection of the 
heavy elements at the base of the convection zone. They deduced from the sound speed profile
a  low value of the tachocline thickness ($h=0.02$) which could be due to the macroscopic coefficient used. 
In this estimate, they did not calibrate their model in Z/X and obtained a result similar to ours (dash dot line in figure \ref{f4}a).

\subsection{Solar composition}

Examining the different observables, it seems that the model using the coefficient
$B_{tz}$ (see Tables \ref{table 1} and \ref{table 2})
 corresponds  today to a good representation of the present Sun. 
The agreement on the sound speed is significantly improved and the observed
photospheric light elements ($^3$He, $^4$He, $^7$Li, $^9$Be) are well reproduced.
In this model, two effects contribute to this improvement; the introduction of 
the mixing at the base of the convection zone, which partly inhibits the microscopic 
diffusion, and the induced increase of the heavy elements at the surface 
and consequently 
in the intermediate region due to this inhibition. 
This model 
treats the tachocline instability as a local phenomenon. 

In Figure \ref{f4}b we show the heavy element profile along the solar 
radius of the reference model and the mixed models (coefficients $B_t$ \& $B_{tz}$) 
with and without the Z/X=0.0245 calibration. We clearly see that without 
the Z/X calibration there 
is a difference in the Z profile, corresponding to an increase of 
the opacity and then to a reduction of the 
disagreement between the seismic sound speed and the theoretical 
one.

In paper I, we had already mentioned that the sound speed difference between the Sun and 
the model may suggest that a slight increase of the heavy element composition of about $5\%$ is required. One 
notes that the results we obtain without specific calibration but keeping the initial heavy element 
composition of the reference model, lead to a photospheric heavy element composition of Z/X=0.0255. This is 
still consistent with the present observational uncertainties (about $10\%$) even if a very recent work seems 
to suggest a possible reduction of the CNO abundances rather than an slight increase (Grevesse \& Sauval 
1998). We have verified this statement by performing a ``pure'' 
microscopic model with Z/X=0.0255. If one compares 
the sound speed profile of model $B_{tz}$ (dot dash line) to the reference model with Z/X=0.0245 (Fig \ref{f1}), it shows 
a better overall agreement with the seismic data. Otherwise, if one compares this model to a ``pure'' 
microscopic model with Z/X=0.0255, one reaches a different conclusion for the radiative part (identical to the comparison of model $B_t$ with the reference model Fig \ref{f1}). 
This intricate process shows that it is not so easy to find the real behaviour of the Sun. If we increase the initial heavy element abundance a little more, we destroy  the agreement of the tachocline region and 
obtain a photospheric composition which is not in as good agreement with the present observations. So 
extensive studies are surely useful 
 to properly disentangle the competing processes.

The remaining part of the sound speed discrepancy located between the nuclear 
core and the convection zone (see Fig \ref{f1}) may come from uncertainties in the equation of 
state or opacity coefficients (Turck-Chi\`eze et al. 1998), or from our simplified treatment of the 
microscopic diffusion. In the present study, we do not consider the different states of ionisation of the 
different species or the radiative acceleration in the calculation of the microscopic diffusivity (Turcotte 
et al. 1998). We also cannot exclude the presence of other mixing 
processes, for example, the meridional circulation or 
the effect of the internal waves
or magnetic field. It seems difficult to think that the residual discrepancy
 comes from insufficient knowledge of the element abundances, as 
an  increase of the heavy element composition will have 
 a dramatic effect in the region of the energy transport transition.

\subsection{Nuclear reaction rates}

The depletion of lithium is of course dependent on the nuclear reaction rates used. In this study,
we have used the recent compilation of the NACRE consortium (Angulo et al. 1999), which recommends 
the nuclear reaction rate $^7$Li(p,$\alpha$)$^4$He of Engstler et al. (1992).
 The astrophysical factor is increased by 30\% in comparison  
with the older value recommended by Caughlan \& Fowler (1988), mainly due 
 to a higher accuracy numerical integration. 
The updated value of $S_{71}(0)=0.0593$ Mev barn increases the lithium depletion during the Sun evolution by 
$\sim 
40\%$ (see Table 3 model $B_{to}$ and $B_t$). This effect is introduced in all the results presented in this study and constitutes an important point which must be  considered in the lithium depletion problem in stars. It is interesting also to notice that the 
reaction rate of the $^7$Li(p,$\gamma$)$^8$B which is not included in our study, mainly because 
this cross section was considered as very small, has been reestimated these
 last years and found 100 times greater than in Caughlan \& Fowler (1988). The impact on the present study 
must be smaller than 10\%, considering the relative importance of the two reactions.

\subsection {Neutrino predictions}

Of course, one issue of improving solar models is to predict more correctly 
the neutrino fluxes produced in the central region of the Sun (Turck-Chi\`eze 1999).
Tables 1 and 2 show the predicted values for all the models considered.
One notices a very small dependence on the macroscopic coefficient used, 
but nevertheless
two kinds of results depending on whether the models are calibrated in Z/X or not.
This difference is connected to the effect of the inhibition of the 
microscopic
 processes which leads to a slight decrease of the neutrino fluxes. 
  Indeed,  we observe, in the calibrated models, a reduction in 
comparison with the reference model of -1.6\% for the Gallex and Sage 
experiments (resp. Hampel et al. 1996 and Abdurashitov et al. 1996), of -5.1\% 
for the Homestake experiment (Cleveland et al. 1998) and -5.6\% for 
SuperKamiokande (Suzuki et al. 1998). It is useful to recall that 
the general effect of the introduction
of the microscopic diffusion in solar model was an increase of about 20\% on 
the $^8$B neutrino flux prediction, so this corresponds to 
25\% of this effect. But,
one may also notice that the effect of inhibition could be exactly 
compensated by the effect of the increase of the heavy elements by $5\%$
(reference model and model $B_{tz}$).

  These effects are too small to solve the neutrino problem. This is hardly a surprise, since the tachocline mixing occurs at the base of the convection zone, hence too remote to have a serious impact on the structure of the nuclear core. Even if this study is not 
crucial for the neutrino fluxes it gives certainly a proper description of the chemical transport of elements
  in the external part of the Sun and definitively excludes a large increase of the neutrino fluxes by this 
kind of processes as was thought
  five years ago. The introduction of a macroscopic diffusivity also destroys the criticisms on the impact of 
the pure microscopic diffusion which was well known to be partly inhibited by some turbulent processes and 
which seems now well under control, at least as far as the extension in the Sun is concerned.  
  We begin to include dynamical effects properly in stellar models and 
   one cannot exclude
 that the cumulative effect of different modifications of the physical inputs (microscopic and 
macroscopic) may lead to smaller neutrino fluxes.
We will not comment, in this paper, on the discrepancy of the sound speed 
in the central region  because this  
 is more difficult to estimate (Turck-Chi\`eze \& Basu 1998, 
 Gonczi et al. 1998) 
 and to interpret there, due to the sensitivity of the sound speed 
 to a lot of competing processes (see paper I). Works are in progress to have  good constraints on the 
core. 

\section{CONCLUSION}

In this paper, we have shown that the presence of mixing in the tachocline layer partly inhibits the 
microscopic diffusion process and improves the agreement with the helioseismic and photospheric abundance 
data. Our results make use of updated helioseismic constraints on the present extent of this layer, and 
they are based on a physical treatment of the tachocline layer (SZ92) which involves only one additional
parameter, namely the Brunt-V\"ais\"al\"a frequency (which we assume constant in our model). The
 solar models thus obtained are quantitatively in good agreement with the observational 
data and our favoured model is model $B_{tz}$ (cf. Tables 2 \& 3). When introducing a time dependent tachocline mixing, it is possible to reach the observed $^7$Li 
photospheric abundance at the present solar age, without destroying $^9$Be or bringing too much $^3$He 
at the surface. This does not necessarily mean that our mixing process is the only one which operates 
in the Sun: others may contribute, such as mixing by gravity waves (Montalban \& Schatzman 1996) or meridional circulation (Zahn 1992, Vauclair \& Richard 1998). But 
the lack of depletion of $^9$Be proves that the mixing occurs in a rather thin layer below 
the convective zone, which is the case in our models. The model which seems the most promising assumes a
Brunt-V\"ais\"al\"a frequency $N=25\mu$Hz; it leads to a slight increase of the heavy element abundance by 
5\% at the surface, compatible with the present uncertainties. A detailed study of the feedback 
of such mixing on the structure of the Sun reveals that its effect is not confined to the base of
the convection zone, but that it spreads into the region below, except if one relaxes the fixed Z/X constraint and let the model adjusts within the heavy element abundance incertitudes (model $B_{tz}$). 

From the results of papers I and II, we could deduce a galactic enrichment of 
$^4$He. We find that the solar $^4$He primordial abundance is $Y_{0}=0.27^
{+0.002}_{-0.001}$ and leads to an enrichment of $10.6^{+2.8}_{-1.3}\%$ from the value of 
$0.244\pm0.002$ deduced by Thuan \& Izotov (1998) from observations of a large sample of low-metallicity  
H II regions in blue compact dwarf galaxies.

This first attempt to calibrate dynamical processes with helioseismology data from SOHO leads to the idea 
that the neutrino problem 
could be seen from a new angle, by considering reliable solar models with macroscopic 
processes, never implemented previously. The process invoked here concerns the 5\% part of the 
external mass and consequently has little impact on neutrino predictions. But, even if the solar 
models are close to the seismic data for $r<0.2 R_{\odot}$, the remaining discrepancies are 
still significant ($\sim$ few $\sigma$) in comparison with seismic uncertainties, and they
suggest that other physical processes, neglected so far, may operate there.

To conclude, this study encourages the introduction of new processes in stellar structure theory, 
beyond the standard  model framework. Of course we are aware of the crudeness of the present
model: a more refined treatment would include the variation with depth of the horizontal turbulent 
diffusivity and of the Brunt-V\"ais\"al\"a frequency, as well as a more precise spin-down law, and it 
would not ignore the magnetic field. 

%\newpage

\newpage

\begin{center}{\large {\bf Table captions}}\end{center}
Definition of the parameters used in Tables \ref{table 0} and \ref{table 1}:\\
$d$: is twice the tachocline thickness $h$, $N$: Brunt-V\"ais\"al\"a frequency, $\alpha$: mixing length 
parameter, Y$_0$, Z$_0$, (Z/X)$_0$: initial helium, heavy element and ratio heavy element on hydrogen, Y$_s$, 
Z$_s$, (Z/X)$_s$: idem for surface compositions, R$_{bzc}$, T$_{bcz}$ are the radius and temperature at the 
base of the convective zone, Y$_c$, Z$_c$, T$_c$, $\rho_c$: central helium, heavy element contents, central 
temperature and density, $^{71}$Ga, $^{37}$Cl, $^8$B respective neutrino predictions for the gallium, the 
chlorine and water detectors. OPAL/A means that we use OPAL96 opacities and Alexander (1994) for low 
temperature.

\begin{table}[!ht]
\begin{center}
\caption[]{Solar models with a time independent macroscopic diffusivity describing the tachocline 
mixing and compared to the reference model, each of them corresponds to different values of the Brunt 
V\"aiss\"al\"a frequency and  to the tachocline thickness.\label{table 0}}\end{center}
\begin{center}
\small
%\vspace{-1cm}
   \begin{tabular}{p{3cm}*{4}{c}}       %% example for 3-column table
    \hline
     Parameters  &  Ref &  $A$ &  $B$ \\
    \hline
     Opacities   & OPAL/A & OPAL/A & OPAL/A\\
     Diffusion    & yes &  yes & yes \\
     Age (Gyr) & 4.6 & 4.6 & 4.6 \\
     $d$ ($r/R_{\odot}$)& - & 0.1 & 0.1\\
     $N$ ($\mu$Hz) & - & 100 & 25 \\
     (Z/X)$_s$ & fixed & fixed & fixed \\
     \\
     $\alpha$  & 1.766 & 1.753 & 1.745\\  
     Y$_0$  & 0.2722 &  0.2703 & 0.2691\\
     Z$_0$  & 0.01959 & 0.01919 & 0.01893 \\
     (Z/X)$_0$ & 0.0277 & 0.0270 & 0.0266 \\
     \\
     Y$_s$  & 0.2427 &  0.2452 & 0.2473 \\
     Z$_s$  & 0.01803 & 0.01797 & 0.01792 \\
     (Z/X)$_s$ & 0.0245 & 0.0245 & 0.0245 \\
     R$_{bzc}$/R$_{\odot}$ & 0.7133 &  0.7145 & 0.7154 \\
     T$_{bzc} \times 10^6$ (K) & 2.190 & 2.181 & 2.174 \\
     \\
     Y$_c$   & 0.6405 & 0.6382 & 0.6367  \\
     Z$_c$   & 0.02094 & 0.02051 & 0.02024 \\
     T$_c \times 10^6$ (K) & 15.71 & 15.68 & 15.66 \\
     $\rho_c$ (g/cm$^3$) & 153.1 & 152.9 & 152.7 \\
     \\
     $\rm ^{71}Ga $ (SNU) & 127.1 & 126.0 & 125.2 \\
     $\rm ^{37}Cl $  (SNU)& 7.04 & 6.84 & 6.71 \\
     $^8$B (10$^6$/cm$^2$/s)  & 4.99 & 4.83 & 4.73 \\
    \hline
   \end{tabular}
  \end{center}
\end{table}

\begin{table}[!ht]
\begin{center}
\caption[]{Solar models with a time dependent macroscopic diffusivity describing the tachocline mixing 
and compared to the reference model, each of them corresponds to different values of the Brunt V\"aiss\"al\"a 
frequency and to 
 the tachocline thickness, the model $B_{tz}$ shows the effect of non calibrated Z/X (see text). 
\label{table 1}}\end{center}
\begin{center}
\small
%\vspace{-1cm}
   \begin{tabular}{p{3cm}*{6}{c}}       %% example for 3-column table
    \hline
     Parameters  &  Ref & $A_t$ & $B_t$ &  $B_{tz}$ & $C_t$ \\
    \hline
     Opacities   & OPAL/A & OPAL & OPAL/A & OPAL/A & OPAL/A\\
     Diffusion    & yes &  yes & yes & yes & yes  \\
     Age (Gyr) & 4.6 & 4.6 & 4.6 & 4.6 & 4.6 \\
     $d$ ($r/R_{\odot}$)& - & 0.1 & 0.1 & 0.1 & 0.05 & \\
     $N$ ($\mu$Hz) & - & 100 & 25 & 25 & 25 \\
     (Z/X)$_s$ & fixed  & fixed & fixed & free & fixed \\
     \\
     $\alpha$  & 1.766 & 1.752 & 1.743 & 1.755 & 1.748\\  
     Y$_0$  & 0.2722 & 0.2701 & 0.2689 & 0.2722 &  0.2695\\
     Z$_0$  & 0.01959 & 0.01914 & 0.01889 &  0.01959 & 0.01903 \\
     (Z/X)$_0$ & 0.0277 & 0.0269 & 0.0265 & 0.0277 & 0.0267 \\
     \\
     Y$_s$  & 0.2427 & 0.2455 & 0.2477 & 0.2510 & 0.2464 \\
     Z$_s$  & 0.01803 & 0.01796 & 0.01791 & 0.01858 & 0.01794\\
     (Z/X)$_s$ & 0.0245 & 0.0245 & 0.0245 & 0.0255 & 0.0245\\
     R$_{bzc}$/R$_{\odot}$ & 0.7133 & 0.7146 & 0.7155 & 0.7141 & 0.715 \\
     T$_{bzc} \times 10^6$ (K) & 2.190 & 2.180 & 2.173 & 2.194 & 2.177\\
     \\
     Y$_c$   & 0.6405 & 0.6379 & 0.6365 & 0.6405 & 0.6372\\
     Z$_c$   & 0.02094 & 0.02047 & 0.02019 & 0.02094 & 0.02034 \\
     T$_c \times 10^6$ (K) & 15.71 & 15.67 & 15.65 & 15.71 & 15.66\\
     $\rho_c$ (g/cm$^3$) & 153.1 & 152.8 & 152.7 & 153.1 & 152.8 \\
     $\rm ^{71}Ga $ (SNU) & 127.1 & 125.9 & 125.1 & 127.1 & 125.5 \\
     $\rm ^{37}Cl $  (SNU)& 7.04 & 6.82 & 6.69 & 7.04 & 6.76 \\
     $^8$B (10$^6$/cm$^2$/s)  & 4.99 & 4.81 & 4.70 & 4.99 & 4.76\\
    \hline
   \end{tabular}
  \end{center}
\end{table}

{\small
\begin{table}[!ht]
\begin{center}
\caption[]{Surface abundance variation of $^3$He/$^4$He during the last 3 Gyr, 
surface abundances of $^4$He and heavy elements Z, and abundance ratio initial/surface for $^7$Li and $^9$Be 
from observations and for solar models at the solar age (subscripts $t$ for time dependent models, $tz$ for 
time dependent model with Z$_0=$Z$_0^{ref}=0.01959$ and $to$ for time dependent model with Caughlan \& Fowler 
(1988) $^7$Li(p,$\alpha$)$^4$He nuclear reaction rate).\label{table 2}}
\vspace{0.1cm}
\begin{tabular}{p{1.8cm}*{10}{c}}
\hline
 & Obs & Ref & $A$ & $B$ & $A_t$ & $B_t$ & $B_{tz}$ & $B_{to}$ & 
$C_t$\\
\hline
 $d$ ($r/R_{\odot}^{.}$)& $\leq0.1$ & - & 0.1 & 0.1 & 0.1 & 0.1 & 0.1 & 0.1 &0.05\\
 $N$ ($\mu$Hz) & -&- & 100 & 25 & 100 & 25 & 25 & 25 & 25 \\
 ($^3$He/$^4$He)$_s$ & max 10\% & 2.28\% & 2.14\% & 2.01\% & 2.11\% & 2.0\% & 2.02\% & 2.0\% & 
2.07\% \\
 $^4$He$_s$ & 0.249$\pm$0.003 &  0.2427 & 0.2452 & 0.2473 & 0.2455 & 0.2477 & 0.2509 & 0.2477 & 
0.2464  \\
 (Z/X)$_s$ & 0.0245$\pm$0.002& 0.0245  & 0.0245 & 0.0245 & 0.0245 & 0.0245 & 0.0255 & 0.0245 & 
0.0245 \\
 $^7$Li$_0$/$^7$Li$_s$ & $\sim$100 & $\sim$6 & $\sim$8 &$\sim$22 & $\sim$12 &$\sim$91 
&$\sim$134 & $\sim$64 & $\sim$89 \\
 $^9$Be$_0$/$^9$Be$_s$ & 1.10$\pm$0.03 & 1.115  & 1.093 & 1.086 & 1.093 & 1.118 & 1.125 & 1.118 
& 1.092 \\
 \hline
\end{tabular}
\end{center}
\vspace{-0.6cm}
\end{table}}

\end{document}